# Band Gap Engineering with Ultralarge Biaxial Strains in Suspended Monolayer MoS$_2$

David Lloyd [1], Xinghui Liu[2], Jason W. Christopher[3], Lauren Cantley[1], Anubhav Wadehra[4], Brian L. Kim[1], Bennett B. Goldberg[3], Anna K. Swan[5], and J. Scott Bunch[1,6*]

[1]Boston University, Department of Mechanical Engineering, Boston, MA 02215 USA

[2]University of Colorado, Department of Mechanical Engineering, Boulder, CO 80309 USA

[3]Department of Physics, Boston University, 590 Commonwealth Avenue, Boston, Massachusetts 02215, United States

[4]Department of Materials and Metallurgy, PEC University of Technology, Chandigarh, India-160012

[5]Department of Electrical and Computer Engineering, Boston University, 590 Commonwealth Avenue, Boston, Massachusetts 02215, United States

[6]Boston University, Division of Materials Science and Engineering, Brookline, MA 02446 USA

*e-mail: bunch@bu.edu, fax: 1-617-353-5866

**Abstract**

We demonstrate the continuous and reversible tuning of the optical band gap of suspended monolayer $MoS_2$ membranes by as much as 500 meV by applying very large biaxial strains. By using chemical vapor deposition (CVD) to grow crystals that are highly impermeable to gas, we are able to apply a pressure difference across suspended membranes to induce biaxial strains. We observe the effect of strain on the energy and intensity of the peaks in the photoluminescence (PL) spectrum, and find a linear tuning rate of the optical band gap of 99 meV/%. This method is then used to study the PL spectra of bilayer and trilayer devices under strain, and to find the shift rates and Grüneisen parameters of two Raman modes in monolayer $MoS_2$. Finally, we use this result to show that we can apply biaxial strains as large as 5.6% across micron sized areas, and report evidence for the strain tuning of higher level optical transitions.

The ability to produce materials of truly nanoscale dimensions has revolutionized the potential for modulating or enhancing the physical properties of semiconductors by mechanical strain[1]. Strain engineering is routinely used in semiconductor manufacturing, with essential electrical components such as the silicon transistor or quantum well laser using strain to improve efficiency and performance[2,3]. Nano-structured materials are particularly suited to this technique, as they are often able to remain elastic when subject to strains many times larger than their bulk counterparts can withstand[4]. For instance, bulk silicon fractures when strained to just 1.2%, whereas silicon nanowires can reach strains of as much as 3.5%[5]. Parameters such as the band gap energy or carrier mobility of a semiconductor, which are often crucial to the electronic or photonic device performance, can be highly sensitive to the application of only small strains. The combination of this sensitivity with the ultra-high strains possible at the nanoscale could lead to an unprecedented ability to modify the electrical or photonic properties of materials in a continuous and reversible manner.

Monolayer $MoS_2$, a 2D atomic crystal, has been shown in both theory[6,7] and experiment[8–12] to be an ideal candidate for strain engineering. It belongs to the class of 2D transition metal dichalcogonides (TMD's), and as a direct-gap semiconductor[13] has received significant interest as a channel material in transistors[14], photovoltaics[15] and photodetection[16] devices. It has a breaking strain of 6-11% as measured by nanoindentation, which approaches its maximum theoretical strain limit[17] and classifies it as an ultra-strength material. Its electronic structure has also proven to be highly sensitive to strain, with experiments showing that the optical band gap reduces by ~50 meV/% for uniaxial strain[8,11], and is predicted to reduce by ~100 meV/% for biaxial strain[18,19]. This reversible modulation of the band gap could be used to make wavelength tunable phototransistors[16], or $MoS_2$ strain sensors that have a sensitivity comparable to their state of the art silicon counterparts[20]. Moreover it has been suggested that strain could also improve the performance of $MoS_2$ transistors[21], or could be used to create broadband light absorbers for energy harvesting[22].

The effect of strain on the band gap of 2D TMD's has been reported in a number of studies[9–12,20,23,24], including uniaxial strains of up to ~4 %[25] and biaxial strains of up to ~3 % produced in highly localized sub-micron areas[26]. Band gap shifts in $MoS_2$ of ~300 meV have been induced by using very large hydrostatic pressures[27], and tensile strain has induced shifts of as much as ~100 meV[11]. However, the combination being both an ultra-strength material and having a band gap highly sensitive to strain imply that a much larger band gap tuning must be possible. By contrast, tensile strain has been used to reduce the band gap by as much as 290 meV in 1D nanowires[28].

In this paper, we use a geometry which allows us to take the first measurements of the Raman mode and band gap shift rates of suspended $MoS_2$ membranes under large biaxial strains, and study single and multilayer samples prepared by both CVD and mechanical exfoliation. We conclude that micron scale CVD grown monolayer $MoS_2$ can be biaxially strained by over 5% resulting in an optical band gap reduction of ~500 meV, or over 25%.

Our geometry exploits the fact that monolayer $MoS_2$, like graphene, is impermeable to all standard gases[29]. By applying a pressure difference across a $MoS_2$ membrane suspended over a cylindrical cavity (Fig. 1a) a bulge is formed, and this deformation produces a biaxial strain at the center of the device. To fabricate our devices, we first suspend $MoS_2$ films over cylindrical micro-cavities etched into a $SiO_x$/ Si substrate by the transfer of CVD grown $MoS_2$ using a PMMA transfer method[30]. Fig 1b shows a typical transfer with a high yield of undamaged suspended devices. We used a novel CVD growth recipe (see supporting information for details) which produces highly impermeable monolayer membranes. With our best growths, a single transfer can produce several hundred suspended monolayer devices which are impermeable to the larger gas species (Fig. S3).

Fig. 1c shows atomic force microscopy (AFM) cross sections of one of these devices under ambient external pressures ($p_{ext} = p_{atm}$) but with increasing internal pressures ($p_{int}$), resulting in increasing center membrane deflections $\delta$. The device can be bulged up ($\delta > 0$) or down ($\delta < 0$) depending on whether the pressure difference across the membrane, $\Delta p = p_{int} - p_{ext}$, is positive or negative. We vary $p_{int}$ by placing the devices in a chamber filled with pressurized $N_2$ gas, which is able to slowly diffuse through the silicon oxide substrate and into the sealed micro-cavities. They are left there for several days until $p_{int}$ equilibrates with the pressure of the $N_2$ gas[29]. After the devices are removed from the chamber, the new $p_{int}$ results in a different $\delta$ and biaxial strain $\varepsilon$ in the center of the device.

Following Hencky's model for circular, pressurized membranes with a negligible bending stiffness[31], the biaxial strain produced at the center of the device can be written as,

$$\varepsilon = \sigma(\nu)\left(\frac{\delta}{a}\right)^2 \quad (1)$$

where $\sigma(\nu)$ is a numerical constant which depends only on Poisson's ratio $\nu$ (see supporting information). For $MoS_2$ we take the value of $\nu = 0.29^{32}$, resulting in $\sigma = 0.709$. This model has been shown to accurately describe graphene membranes in this geometry[33]. We can therefore measure $\varepsilon$ at each $p_{int}$ by using an AFM to find $\delta$ and $a$, and by varying the magnitude of $p_{int}$ we can take optical measurements of the band gap and Raman shifts over a range of known strains.

We first studied the effect of strain on the PL of CVD and mechanically exfoliated monolayer devices, and Fig. 2a shows the PL spectra of a monolayer device over the range of 0 – 2% biaxial strain. We incrementally increased $p_{int}$ up to ~0.75 MPa corresponding to a strain of ~2%, and at successive pressures a PL, Raman and AFM measurement was taken. At higher $p_{int}$, the membranes begin to delaminate from the surface as the force from $\Delta p$ overcomes the adhesion to the substrate[33], which limits the maximum possible strain with $\Delta p > 0$ to ~2%. Membranes in this geometry may slide at the edge of the well under high pressure[34], however we only present data for devices which show no evidence of significant sliding (Fig. S4). For optical measurements we used a 532 nm laser with a spot size of ~ 1 μm in diameter. Our devices were 8 μm in diameter, allowing us to focus the laser spot only on the region of biaxial strain in the center of the device. We observed that the PL peak redshifted with increasing strain and also rapidly decreased in intensity, consistent with previous work[11] and theoretical predictions[7] (Fig. S5).

Each spectrum in Fig. 2a contains peaks resulting from the decay of the neutrally charged A and B excitons at approximately 1.89 eV and 2.05 eV respectively[13], which form when electrons are excited across the direct band gap at the K-point and are bound to holes in the spin-split A and B valence bands. There is also a third peak ($A^-$) centered at 1.86 eV[35] which results from the decay of negatively charged trions which form when additional conduction band electrons bind to A excitons. To determine how all three peaks were affected by strain, we fitted three Voigt functions to each of our PL spectra (Fig. 2b inset), and plotted the peak position of the $A^-$, A and B peaks in Fig. 2b. We found there was no difference in the shift rate between exfoliated and CVD grown devices, and that all three peaks had an approximately equal peak shift rate of -99 ± 6 meV/% which agrees well with theoretical predictions of 105 meV/%[9].

We also took a corresponding Raman spectrum at each $p_{int}$, so we can similarly find the shift rate of the Raman modes with strain (Fig. 2c). The two characteristic peaks of unstrained $MoS_2$, relating to in-plane ($E^1_{2g}$) and out-of-plane ($A_{1g}$) vibrations, are found at 385 $cm^{-1}$ and 405 $cm^{-1}$ respectively. By fitting a Voigt function to each mode, we found that the modes shifted linearly at a rate of -1.7 $cm^{-1}$/% for the $A_{1g}$, and -5.2 $cm^{-1}$/% for the $E^1_{2g}$ which agrees well with theoretical predictions[37] and previous experiments[26]. The differences in these values to those found in hydrostatic pressure studies[36] (in which the $A_{1g}$ mode has the higher shift rate) is likely due to the different type of deformation applied in the two cases. Using the formula[38] $\gamma = [\omega - \omega_0]/[2\varepsilon\omega_0]$, we determine the Grüneisen parameters for the modes to be $\gamma_{E_{2g}^1} = 0.68$ and $\gamma_{A_{1g}} = 0.21$, which are also in good agreement with the values found in earlier studies[39,40]. The position of the $A_{1g}$ peak is known to vary with doping[41], however as this is not the case with the more strain sensitive $E^1_{2g}$ mode, its peak position can be used as a reliable way to measure the internal strain of monolayer $MoS_2$.

Multilayer $MoS_2$ is also a promising material for strain based applications[20], so we used the same procedure to take strain and optical measurements of one bilayer device and five trilayer devices prepared by mechanical exfoliation. For these devices we again observed Raman mode softening for both peaks (Fig. 2c), but with smaller shift rates than were seen for monolayers (see Table S1 in supporting information). The PL spectrum of multilayer $MoS_2$ is distinguished from that of monolayers by the presence of a large additional peak resulting from indirect gap emission[13], referred to as the I peak. The peak positions for the I, A and B peaks are plotted against strain in Fig. 2d. We determined the A peak shift rate to be -91 meV/% for bilayers and -73meV/% for trilayers. The indirect I peak shifted considerably faster than the direct peaks in both bilayers and trilayers, at a rate of -144 meV/% and -110 meV/% respectively.

To overcome the limitation in the magnitude of the applied strain imposed by delamination when *Δp > 0,* we can instead increase $p_{ext}$ of the devices which deflects the membrane downwards. To do this, the devices were placed in a custom-built pressure chamber with a sapphire window which allows optical measurements to be taken at various $p_{ext}$[35]. The internal pressures of the cavities were $p_{int} = 0$, as the devices had been left to equilibrate in a vacuum chamber for several days prior to measurements. By pressurizing the chamber with N$_2$ gas, the greater $-\Delta p$ across the membrane deflects it further downwards and produces an increased biaxial strain at the center of the device.

Fig. 3a shows the PL spectrum as $\Delta p$ is varied from 0 to -1.45 MPa. As before, the A peak redshifts with increasing strain, and also rapidly decreases in intensity. The A peak intensity decreased faster than the B peak, so at the largest strains the peaks were of a comparable intensity. As determined from the energy shift of the A peak, we find that we can shift the band gap in this manner by as much as 500 meV.

At each $p_{ext}$ a Raman spectrum was also taken along with its corresponding PL spectrum. The data is normalized to the silicon peak and plotted in Fig. 3b. We saw the softening of both modes with increasing strain as before, and also observed the strain tuning of the second order 2LA(M) mode (Fig. S7b). Due to the changing deflection of the bulge with pressure, the optical interference between light scattered off the membrane and light reflected off the silicon substrate is altered, which produces the oscillatory behavior in both the peak intensities with increasing pressure. As strains are increased, we observe a dramatic increase in the intensity of the $E^1_{2g}$ mode relative to $A_{1g}$ mode, which is an effect not reported in other studies.

Finally, by assuming the linear relationship we found earlier between the $E^1_{2g}$ Raman mode and biaxial strain holds at the higher strains we are now considering, we use the position of the strain sensitive $E^1_{2g}$ peak to determine the biaxial strain that was produced at each $p_{ext}$ in Fig. 3, and we can therefore determine the strain in our devices by optical measurements only.

The A peak position is plotted against this strain in Fig. 4a, showing that biaxial strains as high as 5.6% can be achieved before membrane rupture. The relationship between the band gap shift and strain remains approximately linear at these high strains with a shift rate of 92 ± 6 meV/%, which is consistent with our earlier findings of 99 ± 6 meV/%.

We also plot the integrated intensities of both peaks (normalized to the silicon peak) against the strain as determined by the $E^1_{2g}$ peak position (Fig. 4b). At the highest strains, there was a three-fold enhancement of the $A_{1g}$ peak, and more than a twenty-fold enhancement of the $E^1_{2g}$ peak. By using the Fresnel equations to model the effects of optical interference on our measurements due to the changing $\delta$ with pressure (Fig. 4b bottom panel and Fig. S6), we find that interference effects cannot explain these enhancements, nor the relative enhancement of $E^1_{2g}$ over $A_{1g}$. We also rule out the changing curvature of our devices when strained as the source of this intensity increase (Fig. S8).

Similar enhancements of the Raman peak intensities have been observed when laser excitation energies are resonant with an electronic transition[42–44]. Here, we maintain a constant laser energy of 2.33 eV, however as biaxial strain induces large changes to the electronic band structure, some transition energies may be moved closer to resonance with the laser excitation energy. We therefore attribute the increase in intensity of both peaks relative to the silicon peak to resonant Raman scattering resulting from the strain tuning of a higher level energy transition to be in resonance with the laser. A likely candidate for this transition is the C exciton at ~2.8 eV[43–45], since the redshift required to lower its energy to resonance with our laser would be ~500 meV, a value consistent with the shift of the A peak at our highest strains. These results demonstrate not only that CVD grown monolayer $MoS_2$ films can withstand the remarkably high strains of 5.6% over micron sized areas, but that higher level optical transitions may also be tuned with strain.

The ability to continuously and reversibly modulate the optical band gap of monolayer $MoS_2$ by up to 25% allows significant control over the optical and electrical properties of the material, an effect which could be used to produce sensitive piezoresistive pressure sensors or broadband light absorbers. We also grew atomically thin membranes by CVD which are highly impermeable to gases and can withstand large pressure differences across them, suggesting that CVD grown $MoS_2$ could be promising as a gas separation membrane. The method used in this work may be extended to study the effects of biaxial strain on other 2D semiconducting materials, and could also be used to determine the effects of very high strains on other strain dependent phenomena, such as magnetism[46], chemical adsorption[47] or piezoelectricity[48].

**Supporting information**

Supporting information includes details of the growth and characterization of membranes, additional Raman shift rate data for multilayer samples, the procedure for device pressurization and optical measurements, gas permeance measurements, a discussion of the Hencky model, evidence of repeatability and the effects of membrane sliding, a comparison to theoretical work of the PL intensity decrease with strain, a discussion of interference effects with additional Raman mode data, and PL mapping data.

**Acknowledgments:**

This work was funded by the National Science Foundation (NSF), grant no. 1054406 (CMMI: CAREER, Atomic Scale Defect Engineering in Graphene Membranes) and grant no. 1411008, a grant to L. Cantley by the NSF Graduate Research Fellowship Program under grant no. DGE-1247312, a grant to J. W. Christopher by the Department of Defense (DoD) Air Force Office of Scientific Research through a National Defense Science and Engineering Graduate (NDSEG) Fellowship, 32 CFR 168a, and a grant to Brian L. Kim by the Boston University Undergraduate Research Opportunities Program (UROP).

**Figure 1**

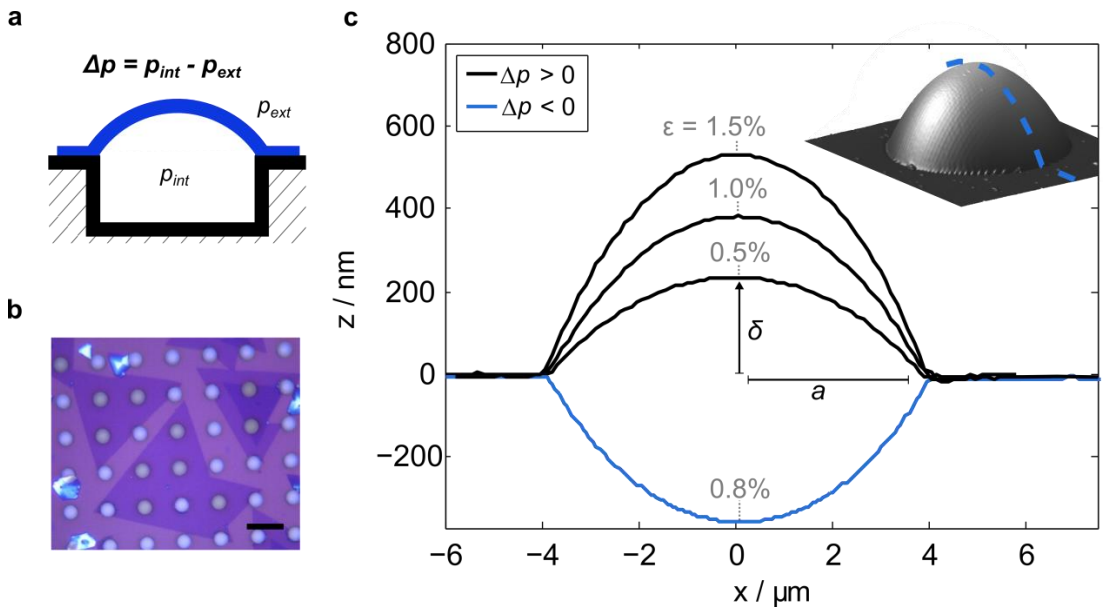

a) Device schematic. b) A typical sample of CVD grown $MoS_2$ membranes suspended over cylindrical cavities after transfer (scale bar is 20 μm). c) An AFM cross section of a device at various $p_{int}$, resulting in different biaxial strains at the center of the device. Devices can be bulged up or down depending on whether $\Delta p$ is positive or negative.

**Figure 2**

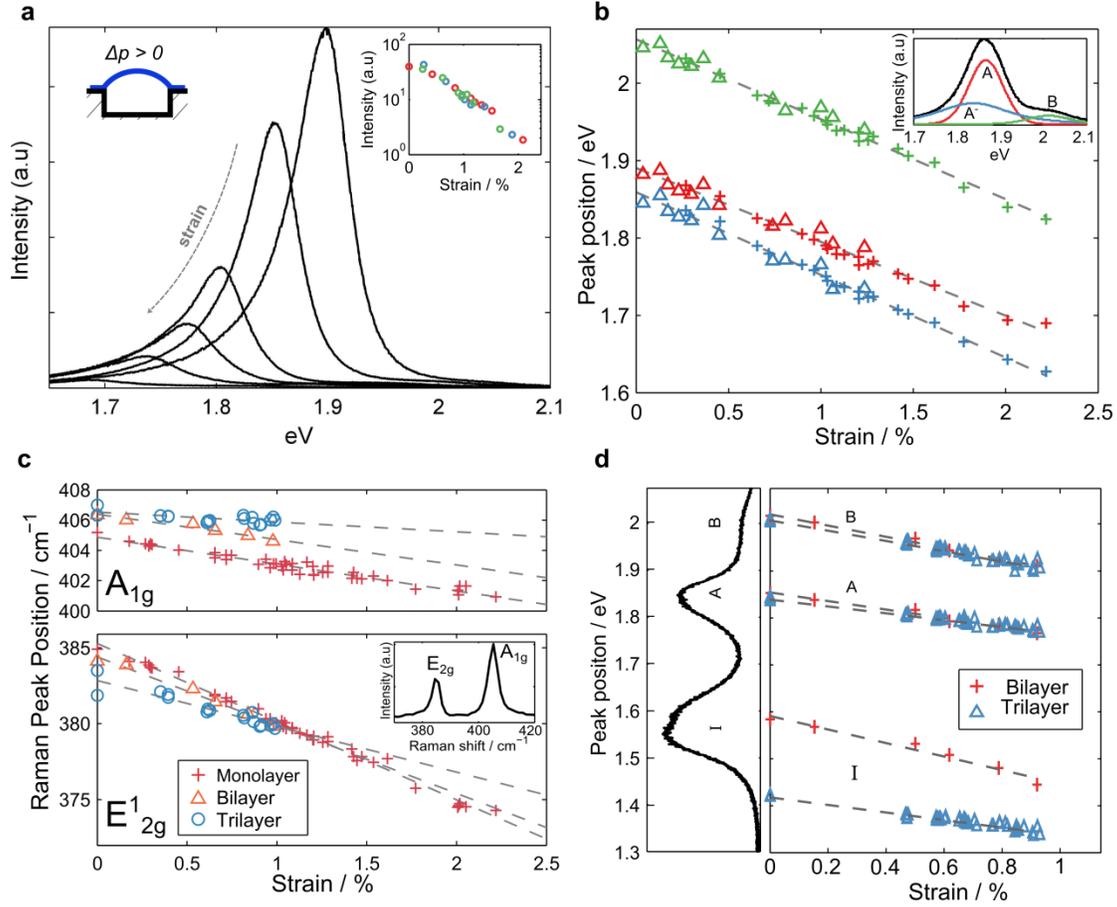

a) The PL spectra for monolayer $MoS_2$ at different biaxial strains corresponding to different $p_{int}$, and the relationship between strain and A peak intensity (inset). Intensities are normalized to the $A_{1g}$ Raman peak. b) The peak positions of the A (red), $A^-$ (blue) and B (green) excitons as a function of biaxial strain for CVD (crosses) and exfoliated (triangles) monolayer devices. The peaks were fitted using three Voigt functions. c) The $E^1_{2g}$ and $A_{1g}$ Raman modes for unstrained $MoS_2$ (inset) and peak positions as a function of biaxial strain for different membrane thicknesses. d) A bilayer PL spectrum, and the peak positions of the A, B and indirect I peak as a function of biaxial strain for exfoliated bilayer and trilayer devices.

**Figure 3**

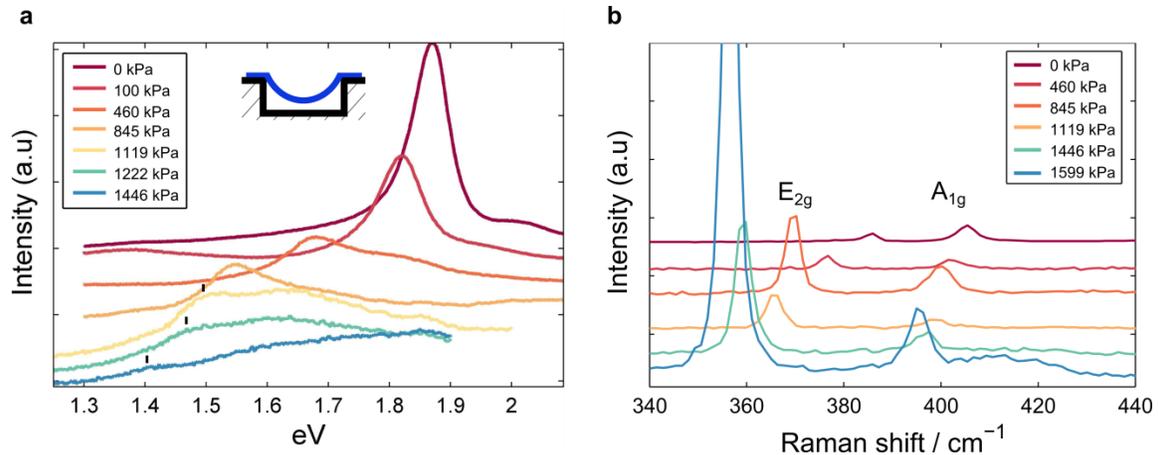

*In-situ* measurements of a) PL spectra for a monolayer device (scaled for comparison with ticks marking A peak position), with the largest pressure difference representing ~5 % strain. b) Raman spectra at increasing chamber pressures. Labels refer to the negative pressure difference $-\Delta p$ across the membrane, and Raman peaks are normalized to the silicon peak intensity.

**Figure 4**

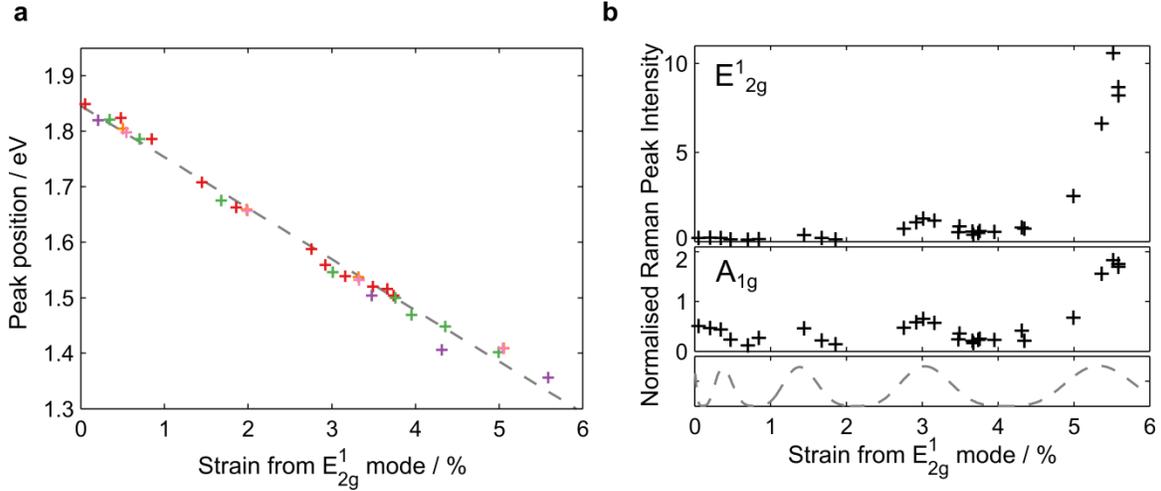

a) The A peak position of the PL spectrum plotted against the strain as determined from the $E^1_{2g}$ peak shift. In this case we fitted a single curve to the A peak feature, as the large decrease in PL intensity meant that the individual A and $A^-$ peak contributions could not be resolved. Different colors represent different devices. b) The integrated intensities of the $E^1_{2g}$ and $A_{1g}$ modes normalized to the silicon peak and plotted against strain. The expected intensity modulation due to interference is also plotted for comparison.

**For TOC only:**

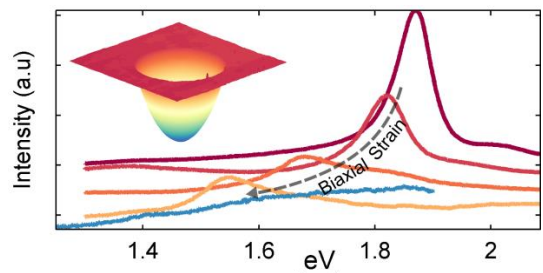

# Supporting Information

# Band Gap Engineering with Ultralarge Biaxial Strains in Suspended Monolayer MoS$_2$

David Lloyd [1], Xinghui Liu[2], Jason W. Christopher[3], Lauren Cantley[1], Anubhav Wadehra[4], Brian L. Kim[1], Bennett B. Goldberg[3], Anna K. Swan[5], and J. Scott Bunch[1,6*]

[1]Boston University, Department of Mechanical Engineering, Boston, MA 02215 USA

[2]University of Colorado, Department of Mechanical Engineering, Boulder, CO 80309 USA

[3]Department of Physics, Boston University, 590 Commonwealth Avenue, Boston, Massachusetts 02215, United States

[4]Department of Materials and Metallurgy, PEC University of Technology, Chandigarh, India-160012

[5]Department of Electrical and Computer Engineering, Boston University, 590 Commonwealth Avenue, Boston, Massachusetts 02215, United States

[6]Boston University, Division of Materials Science and Engineering, Brookline, MA 02446 USA

*e-mail: bunch@bu.edu, fax: 1-617-353-5866

# 1. CVD growth

To grow highly impermeable monolayer $MoS_2$, we use a modified version of the growth method described in Ref 1. A powder source of $MoS_2$ is placed in the center of a furnace, and a $SiO_x$ substrate is placed in a cooler region downstream. The system is pumped down to 10 mTorr to remove any contaminating gases after which we flow 60 sccm Ar as a carrier gas, plus 0.1 sccm of $O_2$ and 1 sccm of $H_2$ gas. The furnace is heated to 900 $^o$C and held at that temperature for 15 minutes after which it is left to cool naturally to room temperature.

The process described in Ref. 1 depends on the sublimation of $MoS_2$ at the hottest part of the furnace which is carried downstream and condenses on the substrate in a cooler region. We found that the yield could be considerably improved by the addition of small amounts of oxygen and hydrogen. This led to large monolayer coverage with triangular sheets with a side length as large as 150 μm. The likely mechanism for this growth is that the $O_2$ reacts to form either $MoO_2$ or $MoO_3$, thus liberating 2S. These molecules then flow downstream to react on the surface of the substrate. This method is therefore analogous to several other methods in the literature[2], in which a molybdenum oxide and sulfur powder precursors are used. However, we found our method to give a considerably larger and more reliable yield of impermeable $MoS_2$ membranes.

# 2. Optical measurements and multilayer characterization

For PL and Raman measurements, we used 600 l/mm and 2400 l/mm gratings respectively. Laser powers were kept below 10 μW/m$^2$ to avoid heating effects, which can modify the PL peak positions or open permeable holes in the membranes. We found our unstrained suspended $MoS_2$ membranes had a high intensity A peak at ~1.88 eV in the PL spectrum, and a Raman mode peak separation of ~19 cm$^{-1}$ which confirms that our samples were of a monolayer thickness.

PL and Raman measurements were also used to confirm the thickness of our bilayer and trilayer samples (inset Fig. S1). Our multilayer samples were prepared by mechanical exfoliation so the stacking orientation between the layers was not controlled; however the Raman peak separations for our bilayer and trilayer samples were 21.7 cm$^{-1}$ and 23.7 cm$^{-1}$ respectively, which are consistent with exfoliated samples in previous reports[3].

| Thickness | $E^1_{2g}$ mode shift rate (cm$^{-1}$ / %) | $A_{1g}$ mode shift rate (cm$^{-1}$ / %) |
|---|---|---|
| Monolayer | -5.2 | -1.7 |
| Bilayer | -4.2 | -1.3 |
| Trilayer | -3.0 | -0.7 |

Table S1. Raman mode shift rates for each membrane thickness.

A comparison of the strain dependencies of the Raman modes that we observed in Fig. 2c of the main text is presented in Table S1. These results show that both modes are less strain sensitive with increasing membrane thickness, an effect which was also observed in Ref 4. We also plot the ratio of the integrated intensities of the two Raman modes ($E^1_{2g}/A_{1g}$) in Fig. S1. All three thicknesses have approximately the same $E^1_{2g}/A_{1g}$ at zero strain, and each show some small increase in this ratio with strain.

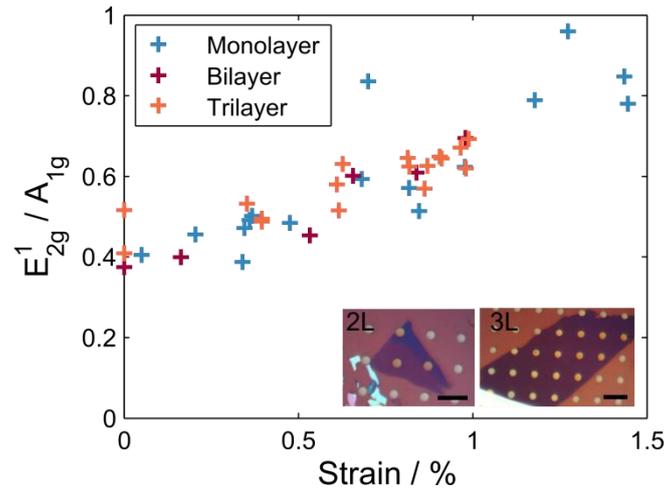

Figure S1. The ratio of integrated intensities of the $E^1_{2g}$ to $A_{1g}$ Raman modes for different membrane thicknesses. Scale bars are 15 μm.

# 3. Pressurization

**a**

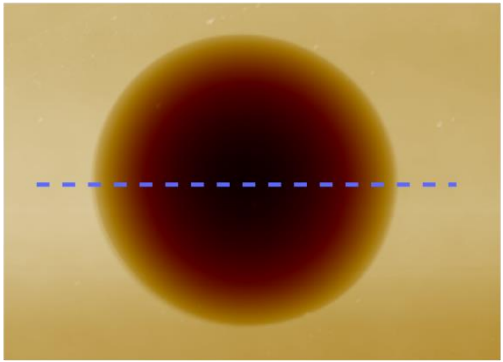 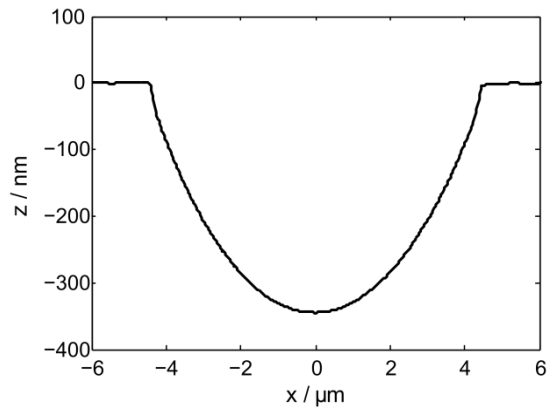

**b**

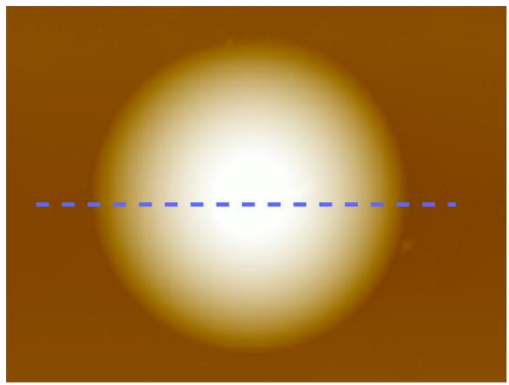 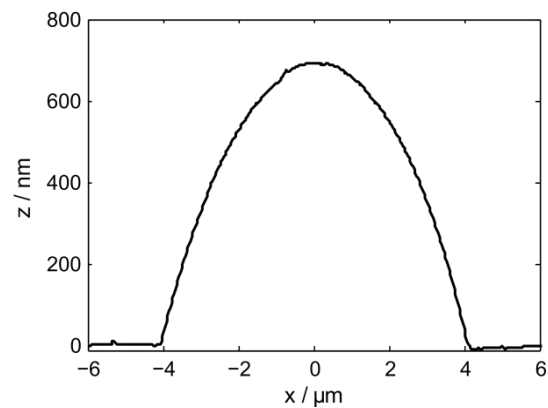

Figure S2. AFM images and cross sections of a) a device with $p_{int} = 0$ and b) a device with $p_{int} \sim 0.75$ MPa. In both a) and b) $p_{ext} = 1$ atm.

The devices used to collect data in Fig. 3 of the main text had $p_{int} = 0$ by leaving them to equilibrate in a vacuum chamber for several days. The only exception to this was for the data points at $\Delta p = 0$, which were taken by inflating the devices with gas until $p_{int} = 1$ atm and the membranes were unstrained. Fig. S2a shows an AFM image and cross section of such a device at $p_{ext} = 1$ atm and $p_{int} = 0$ so that $\Delta p < 0$. This image was taken immediately after the device had been exposed to $p_{ext} \sim 1380$ kPa for several hours, demonstrating that very little gas leaked into the device cavities during that time. Fig. S2b is an image of one of the CVD devices used to take Raman and PL data in Fig. 2 of the main text where $p_{ext} = 1$ atm and $p_{int} \sim 0.75$ MPa so that $\Delta p > 0$.

# 4. Gas permeance

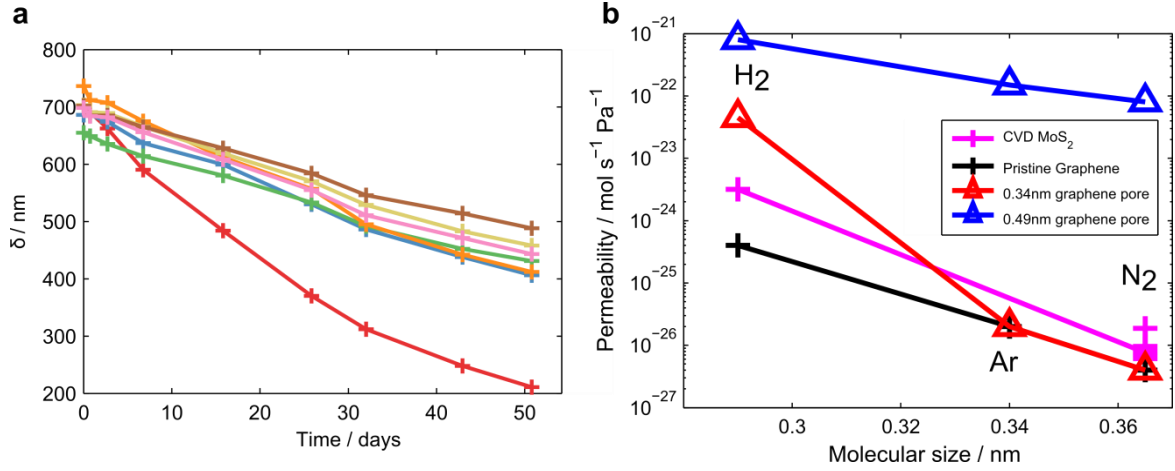

Figure S3. a) Seven different CVD MoS$_2$ devices pressurized with nitrogen and deflating over time. b) Comparison of CVD MoS$_2$ with pristine and etched graphene.

The permeability of our CVD MoS$_2$ membranes can be quantified by measuring the rate at which gas leaks out of the pressurized micro-cavities. To do this, we follow the method used in Ref 5. Briefly, we assume that the leak rate of number of moles of gas *n* can be written in terms of permeability *k* and the pressure difference across the membrane *Δp* as,

$$\frac{dn}{dt} = -k\Delta p \qquad (1)$$

Using the ideal gas law, we can determine *dn/dt* and *Δp* in terms of the maximum deflection of the membrane bulge *δ*. In this way, we can determine *k* by taking AFM measurements of *δ* over time as devices pressurized with various gases deflate. Fig. S3a shows how *δ* changes as seven CVD MoS$_2$ devices initially pressurized with N$_2$ gas deflate over time, and shows that it takes several months for most devices to fully deflate. We can compare the permeability k for these membranes with graphene devices in the same geometry for gases of varying molecular size in Fig. S2b. The graphene data was taken from Ref 6. For both nitrogen and hydrogen, the permeability of the MoS$_2$ is higher than for pristine graphene, however orders of magnitude less than a graphene membrane with an estimated 0.49 nm pore. We can therefore conclude that our CVD membranes must be completely free of permeable nanometer scale vacancies.

# 5. Hencky's solution for the uniform lateral loading of circular membranes.

The mechanics of a uniform pressure on an atomically thin membrane over a cylindrical cavity is described in detail elsewhere[7,8], and this discussion closely follows Fichter (1997). Briefly, by assuming the uniform lateral loading of the membrane, the governing equations can be written in terms of radius $r$ and pressure difference $\Delta p$ as,

$$\sigma_r \frac{dw}{dr} = -\frac{\Delta p r}{2} \quad (2)$$

$$\sigma_\theta = \frac{d}{dr}(r\sigma_r) \quad (3)$$

with the linear stress-strain relationship,

$$\sigma_\theta - v\sigma_r = Et\varepsilon_\theta \quad (4)$$

The radial stress, $\sigma_r$ and the vertical deflection $w$ can be expanded as an infinite series of even powers of radius $r$, and their exact form can be determined by solving for S2 and S3. The final expression for biaxial strain is then written as,

$$\varepsilon_b = \left(\frac{\delta}{a}\right)^2 \frac{b_0(v)(1-v)K(v)^{2/3}}{4} \quad (5)$$

in terms of two numerical constants $b_0$ and $K$ which both depend only on Poisson's ratio $v$. For $v = 0.29$, $K = 3.54$ and $b_0 = 1.72$, which gives the value of $\sigma = 0.709$ used in the main text.

# 6. Sliding and Repeatability

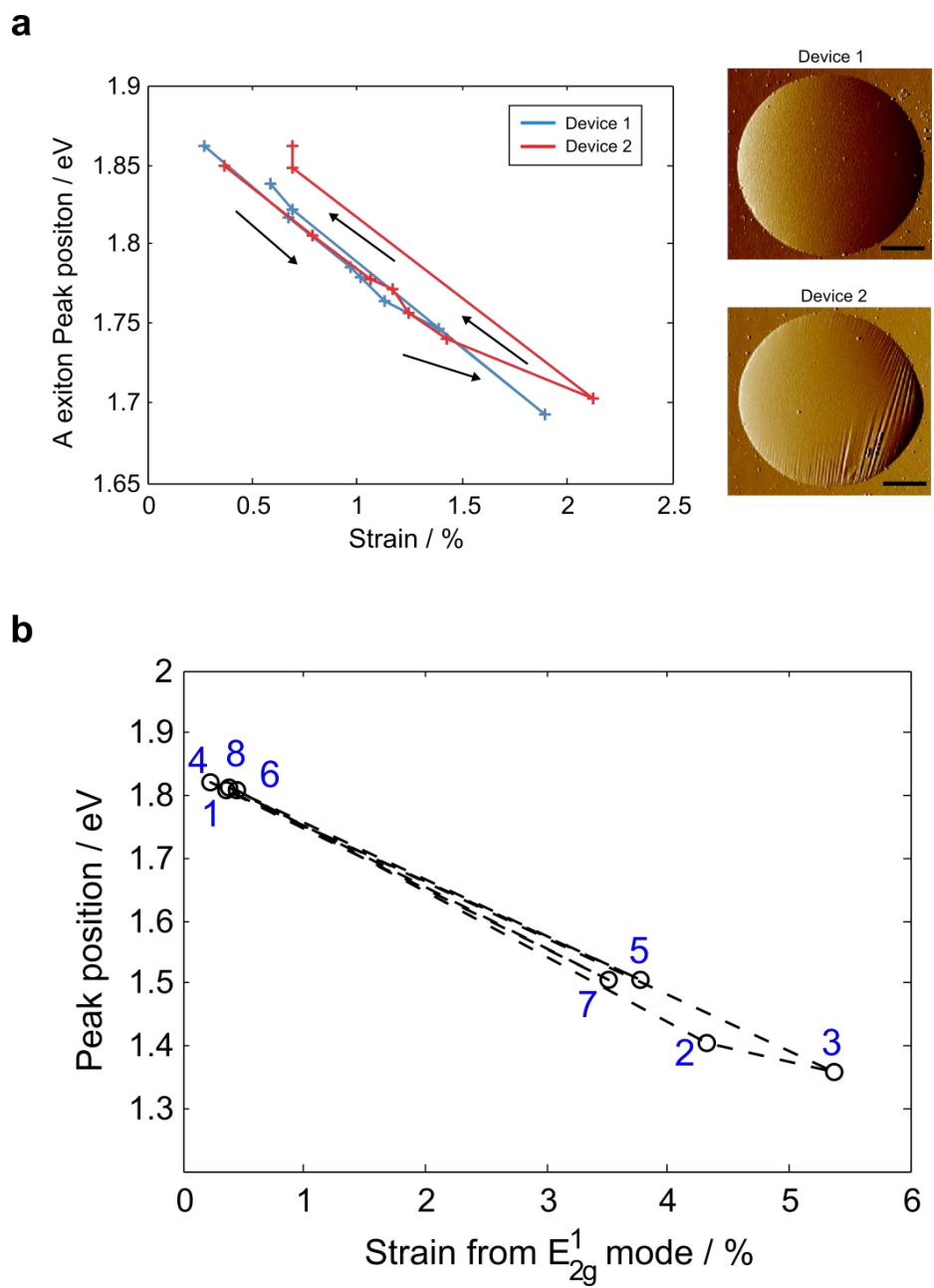

Figure S4. a) An example of the peak shift in two monolayer devices in which $p_{int}$ was increased then decreased, with AFM images of both devices after pressure cycling (scale bars are 2 μm). Device 2 shows evidence of slipping. b) The repeatability of subjecting a device to high strain. Measurements were taken in the sequence indicated.

Under high strains our devices may be forced to slide over the substrate, an effect which has been observed in graphene[9]. This sliding would allow the membrane deflection $\delta$ to increase, and thus cause us to over-estimate the strain from our AFM measurements. To see if sliding has occurred, we plot the relationship between strain and the *A* peak position during the initial increase in internal pressures followed by the deflation of devices (Fig. S4). Device 1 shows little hysteresis, however Device 2 shows evidence of significant sliding. This can further be confirmed by the AFM images of the devices after deflating, with Device 2 showing wrinkling which was not previously present. To avoid any influence of this effect on our data presented in Fig. 2 of the main text, we only used data taken from devices which showed none of these signs of sliding.

# 7. Decrease in the *A* exciton intensity - comparison to theoretical predictions.

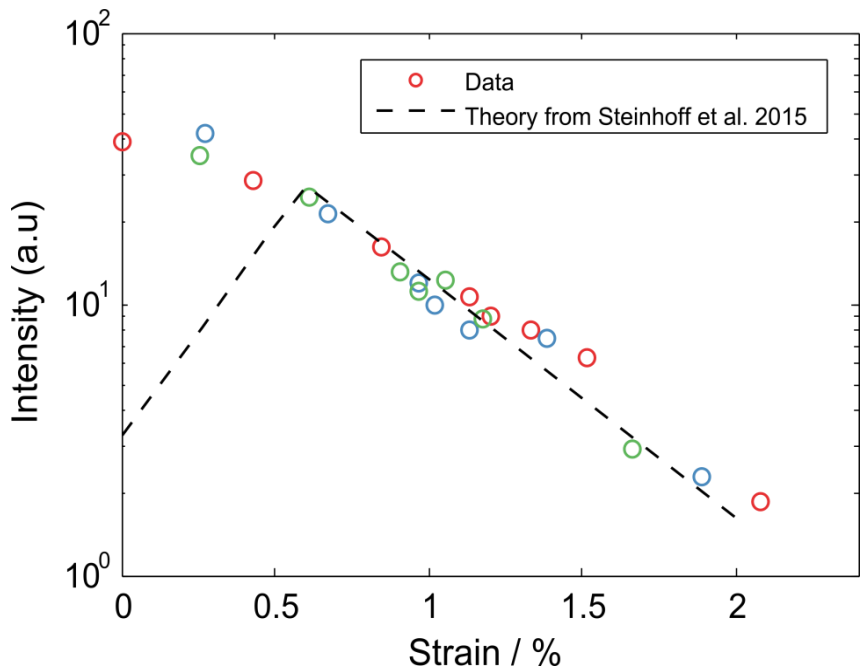

Figure S5. Comparison of the change in *A* peak exciton intensity with strain to a theoretical prediction. The intensity of our data has been scaled for comparison with the theory.

The exponential decrease in *A* peak intensity shown in Fig 2a inset of the main text compares well to the theory described in Steinhoff et al. at large strains (Fig. S5). At strains below 0.5 % however, they predicted that the PL intensity would increase with strain, caused by changes in conduction band minima at the $\Sigma$ point of the conduction band. We did not observe such an enhancement over this range, however the difference may be due to different estimates of the doping level in the theory and the true doping level in our devices.

# 8. Interference effects

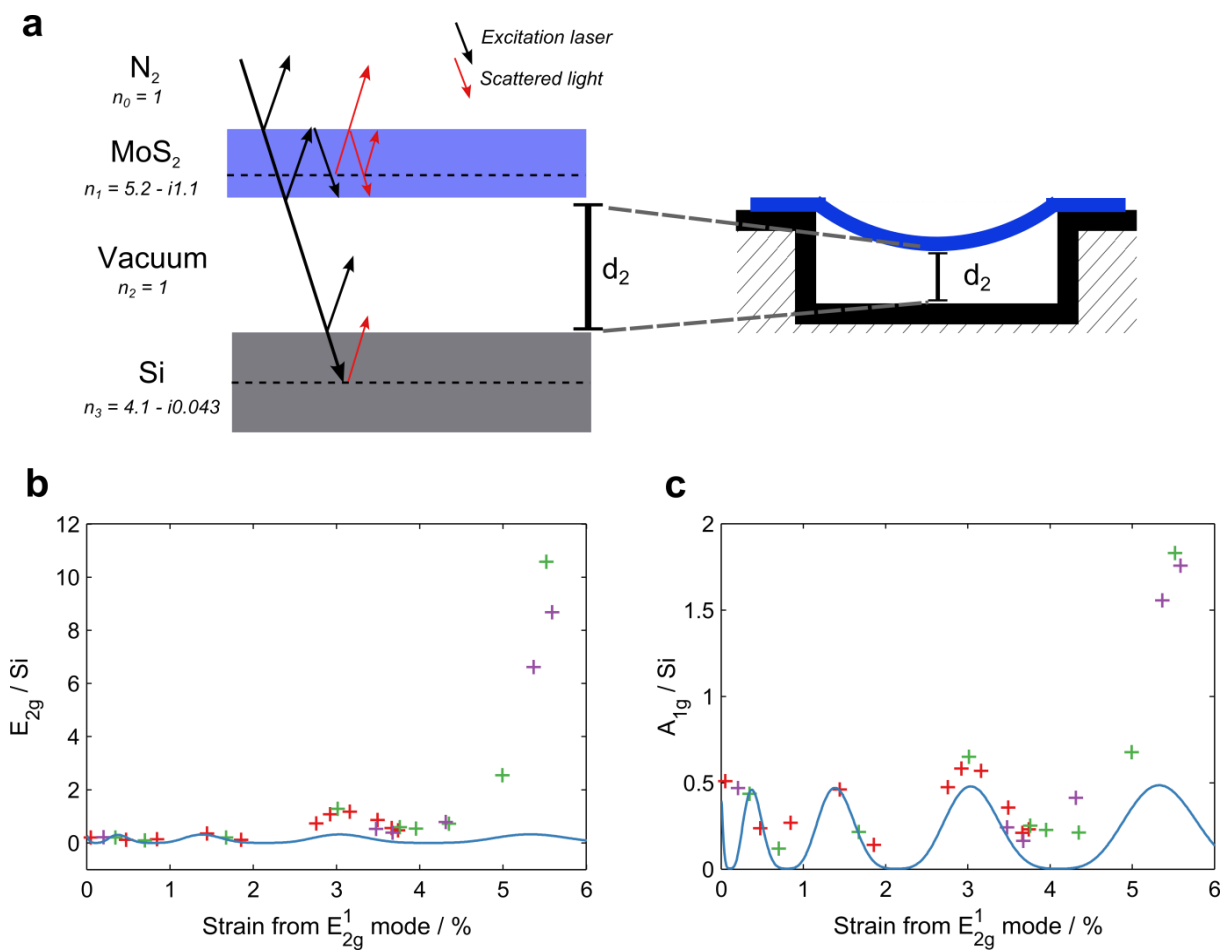

Figure S6. a) A ray diagram of incident and scattered light. The effective power of the excitation laser (black) is the sum of the incident beam with its reflected beams. The intensities of Raman scattered light (red) also depend on the sum of reflected rays, and rays scattered at different phases within the $MoS_2$ or Si. The effect of this interference for each frequency of light depends on the distance between membrane and substrate, $d_2$, which changes as the devices are strained. b) & c) The intensities of the $E^1_{2g}$ and $A_{1g}$ peaks relative to the Si peak. We compare our data (crosses) to the interference model (blue line).

The interference between light scattered of the $MoS_2$ membrane and the silicon at the bottom of the well may affect the relative intensities of the $A_{1g}$ and $E^1_{2g}$ modes (Fig S6a). To rule this out as the cause of our observed changes to Raman mode intensities, we closely follow a model developed in other works[10,11], originally used to determine the effect of the substrate thickness on the Raman mode intensities. In our case, instead of a layer of $SiO_2$, we have a vacuum cavity of distance $d_2$ which changes as the device bulges down under high pressures.

Using the Fresnel equations, we can deduce the change in intensity of each $MoS_2$ Raman mode relative to the Si peak, as $d_2$ decreases from its unstrained value of 1.5 μm (equivalent to the depth of the well) with increasing strain. We deduce the value of $d_2$ by using the Hencky model described in the main text, which can be used to convert strain we determined from the $E^1_{2g}$ peak position, to a membrane deflection $\delta$. $d_2$ is then equal to the difference between $\delta$ and the well depth. We also account for the strain induced shift of each mode in these calculations, which makes the wavelength of the scattered light also dependent on strain.

The model is plotted against our data in Fig. S6b & Fig. S6c, and the intensity of the model curve is scaled in order to be in coincidence with our data at low strains. We find both the $E^1_{2g}$ and $A_{1g}$ peak intensities increase beyond what is expected from the interference model at high strains. We therefore conclude that the observed increase in both Raman mode intensities at high strain was not due to the effect of interference, but was rather an intrinsic property of the material under strain. Similarly, the interference model does not account for why the ratio $E^1_{2g}/A_{1g}$ increases so dramatically. As the Raman modes are so similar in energy, interference effects should cause less than a 10% change in this ratio, and so we conclude that this is also strain induced effect.

## 9. Additional Raman data

We plot the ratio of the Raman modes $E^1_{2g}/A_{1g}$ against strain (Fig. S7a), and at the highest strains there is a ten-fold enhancement of this ratio. To further confirm that the changes in this ratio were not due to the changing distance $d_2$, we also plot the data taken from our bulged up devices. Despite having a different well depth, and a $d_2$ which was increasing with strain (rather than decreasing), the data from the bulged up devices shows a similar trend to that of the bulged down devices. This is further evidence that this ratio change was not caused by interference.

In Fig. S7b we show a zoomed in version of the data presented in Fig. 3b of the main text, in order to show the shift of second order peaks with strain. In the unstrained membrane, this feature is centered around 455 cm$^{-1}$, and we therefore identify it as the 2LA(M) feature[12]. This feature is very strain sensitive, and shifts down to ~415 at 5.5% strain. Like the other peaks its intensity increases with strain, which is further evidence that our laser line is crossing a resonance.

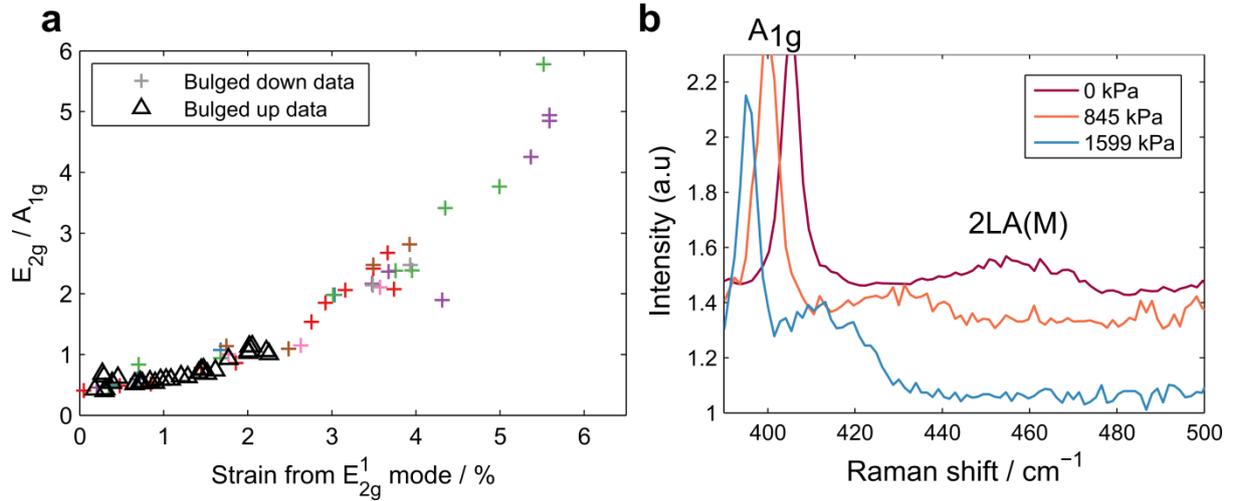

Figure S7. a) The Raman intensity ratio $E^1_{2g}/A_{1g}$ for bulged up (black triangles) and bulged down (colored crosses) devices. b) A zoomed in version of Fig. 3b of the main text to highlight the 2LA(M) mode.

Finally, we did a line scan of a device with $\Delta p$ = 1599 kPa, corresponding to a biaxial strain at the center of the device of ~5.6 % (Fig S8a). As strain increases in our devices, the curvature of the membranes also increases. As our laser spot has a finite size, this change in curvature would change the angle between the incident light and the membrane. To rule this out as a cause of the change in Raman mode intensities, we plot the intensities of the Raman modes as a function of distance $x$ across the device. If changes in angle between laser light and membrane were causing increases in Raman intensity, we would expect the largest change to occur at the edge of the membrane, where the angle change would be the most. However, Fig. S8c and Fig. S8d show that this is not the case. Intensities of both Raman modes and the ratio $E^1_{2g}/A_{1g}$ are both largest in the center of the device, where the biaxial strain is the most and the membrane is closest to being flat. We therefore rule out the device curvature as the cause of these intensity changes.

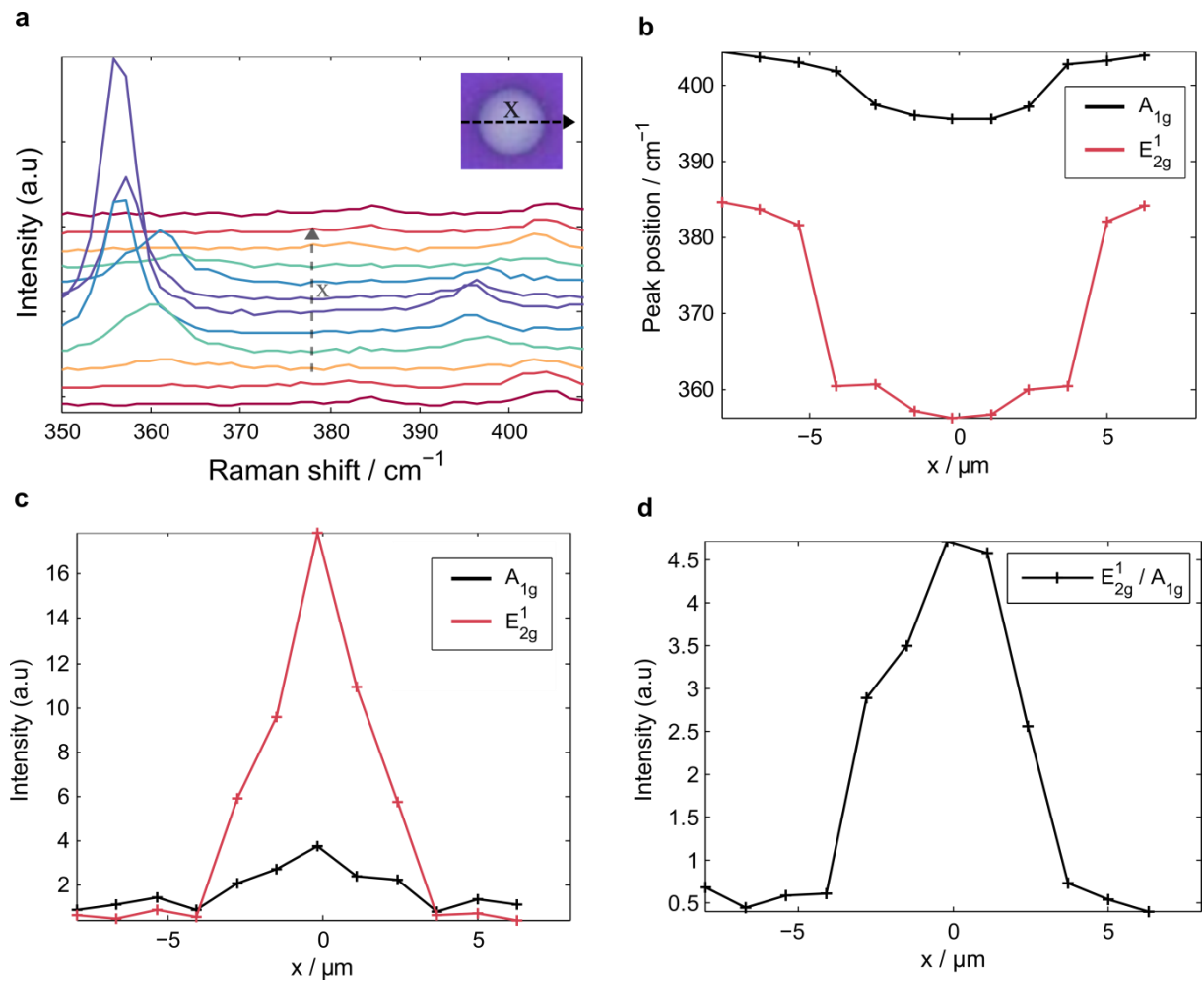

Figure S8. a) Line scan of Raman modes across a device. b) Peak positions and c) peak intensities of the Raman modes (normalized to the Si peak) across the device. d) The ratio $E^1_{2g} / A_{1g}$ of the two Raman mode intensities across the device.

## 10. Additional PL data

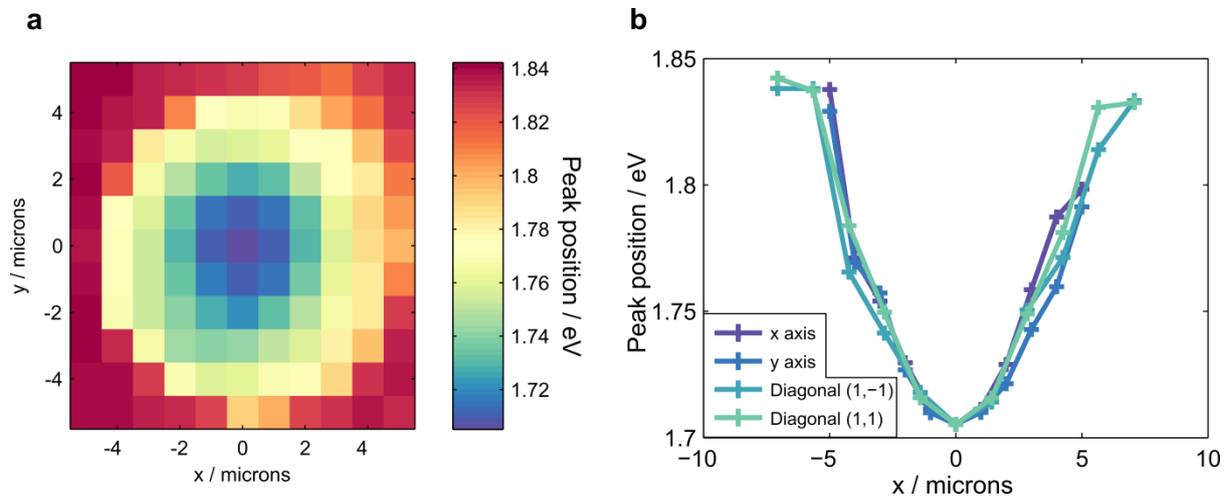

Figure S9. a) A PL map of a device with $p_{int}$ ~ 0.75 MPa and $p_{ext}$ = 1 atm, with colors representing the A peak position. b) Line cuts through the center of the device along the x, y, (1,1) and (1,-1) directions.

The inhomogeneous strain field we apply across our devices should produce a spatially varying optical band gap. To confirm this, we took a PL map of a strained device and plotted the peak position of the A peak at each pixel in Fig. S9a. We also plot line cuts through the center of the device along the axial and diagonal directions (Fig. S9b). The band gap is redshifted to ~1.7 eV at the center of the device where the membrane is subject to a pure biaxial strain. Around the edge of the device the strain becomes approximately uniaxial along the radial direction, which results in a lower band gap shift due to the smaller uniaxial band gap tuning rate in $MoS_2$. These results demonstrate that our device geometry produces an energy gradient which could allow excitons produced around the edge of the device to be funneled towards the lower energy region at the center of the device[13].